\documentclass[a4paper,11pt]{article}
\usepackage{pos}

\newcommand{\GeV}{\rm GeV}
\newcommand{\MeV}{\rm MeV}
\newcommand{\keV}{\rm keV}

\title{
Flavor Probes of Axion Dark Matter
}

\author*[a,b]{Robert Ziegler}

\affiliation[a]{Institut f\"ur Theoretische Teilchenphysik, Karlsruhe Institute of Technology, \\ Wolfgang-Gaede-Str. 1, 76131 Karlsruhe, Germany}

\affiliation[b]{Physikalisches Institut, 
Albert-Ludwigs-Universit\"at Freiburg \\
Hermann-Herder-Str. 3, 79104 Freiburg, Germany}

\emailAdd{robert.ziegler@kit.edu}

\abstract{Standard Model extensions with light axions are well-motivated by the observed Dark Matter abundance and the Peccei-Quinn solution to the Strong CP Problem. In general such axions can have large flavor-violating couplings to SM fermions, which naturally arise in scenarios where the Peccei-Quinn symmetry also explains the hierarchical pattern of fermion masses and mixings. I will discuss how these couplings allow for efficient axion production from the decays of SM particles, giving the opportunity to probe flavored axion Dark Matter with precision flavor experiments, astrophysics and cosmology.}

\FullConference{%
  8th Symposium on Prospects in the Physics of Discrete Symmetries (DISCRETE 2022)\\
  7-11 November, 2022\\
  Baden-Baden, Germany
}

\begin{document}
\maketitle

\section{Introduction}
Axions are excellent Dark Matter (DM) candidates, since they are pseudo-Goldstone bosons of a Peccei-Quinn (PQ) symmetry that is spontaneously broken at high energies. Being approximate Goldstone bosons implies that axions are very light, with masses much below the electroweak scale, while the PQ breaking scale is typically much larger than the electroweak scale, and suppresses all interactions of the axion with Standard Model (SM) particles. For sufficiently small axion masses and sufficiently large PQ breaking scales,  axions can therefore easily be  stable on cosmological scales, with a lifetime roughly given by
\begin{align}
1/\Gamma (a \to \gamma \gamma) \simeq   10^{12} {\rm \, yrs} \left( \frac{f_a}{10^9 \GeV}\right)^2 \left( \frac{ \keV}{m_a}\right)^3 \, , 
\end{align} 
where $m_a$ is the axion mass and $f_a$ the axion decay constant, which is of the order of the PQ breaking scale. The required hierarchy between $m_a$ and $f_a$  is naturally realized for the QCD axion, which also solves the Strong CP Problem, as the mass is inversely proportional to the PQ breaking scale
\begin{align}
m_a = 5.7 {\rm \, meV} \left( \frac{10^9 \GeV}{f_a} \right) \, .
\end{align} 
At scales much below the PQ breaking scale, the most general axion couplings to the SM can be described by an effective field theory (EFT) that includes anomalous couplings to gauge bosons and derivative couplings to fermions~\cite{Georgi}: 
\begin{align}
{\cal L}_{\rm eff} =  \frac{a}{f_a} \frac{\alpha_s}{8 \pi} G \tilde{G} + C_{a \gamma} \frac{a}{f_a} \frac{\alpha_{\rm em}}{8 \pi} F \tilde{F} + \frac{\partial_\mu a}{2 f_a} \sum_i C_i  \overline{f}_i \gamma^\mu  \gamma_5 f_i + \frac{\partial_\mu a}{2 f_a} \sum_{i \ne j}  \overline{f}_i \gamma^\mu \left( C_{ij}^V + C_{ij}^A \gamma_5 \right) f_j \, , 
\end{align}
where I omitted axion couplings to electroweak gauge bosons, which are not phenomenologically relevant for DM axions. Usually only flavor-diagonal couplings $C_i$ are considered, but flavor-violating couplings\footnote{These are hermitian matrices in flavor space, cf. also Eq.~\eqref{CVA}.} $C_{ij}^{V,A}$ are part of the most general axion EFT. They are naturally present in well-motivated UV scenarios and allow for a plethora of new axion couplings to SM particles. While the inclusion of these couplings proliferates the axion parameter space, it also enriches axion phenomenology, as it allows to produce axions via the decay of SM leptons, mesons and baryons such as $\mu \to e a, K \to \pi a, B \to K a, \Lambda \to na$. These processes allow to probe axions  with precision flavor factories, extreme astrophysical environments such as core-collapse supernovae and the early universe. Before I discuss these possibilities in detail, I briefly review the UV origin of flavor-violating axion couplings, and discuss UV scenarios where these couplings naturally arise by identifying the PQ symmetry with a subgroup of a flavor symmetry that addresses Yukawa hierarchies.   
\section{ Motivation}
In general axion couplings to SM fermions arise by rotating the PQ charge matrices to the fermion mass basis, e.g.
\begin{align}
C^{V,A}_{d_i d_j} \propto \left( V_{d_L}^\dagger {\rm PQ}_q V_{d_L}\right)_{ij} \pm \left( V_{d_R}^\dagger {\rm PQ}_d V_{d_R}\right)_{ij} \, , 
\label{CVA}
\end{align}
where $PQ_{q,d}$ denote the PQ charges of SM down quarks and $V_{d_{L,R}}$ are the unitary matrixes that diagonalizes the down-quark mass matrix as $V_{d_L}^\dagger M_d V_{d_R} = M_d^{\rm diag}$. This structure implies that flavor-violating couplings arise whenever SM fermions carry PQ charges that are not aligned to SM Yukawas, i.e. they are not diagonal in the same basis 
\begin{align}
[ {\rm PQ}_q, Y_d Y_d^\dagger ]  \ne 0 \, ,   \qquad {\rm or}   \qquad    [{\rm PQ}_d, Y_d^\dagger Y_d ]  \ne 0 \, .
\end{align}
In the most common QCD axion benchmark models flavor alignment is realized, as PQ charges of SM fermions either vanish (KSVZ models~\cite{KSVZ1,KSVZ2}) or are taken to be flavor-universal  (standard DFSZ models~\cite{DFSZ1,DFSZ2}). However, in a generic situation where PQ charges constitute a new source of flavor violation not aligned to SM Yukawas, flavor-violating axion couplings do arise, for example in variant DFSZ models with flavor non-universal charges~\cite{VariantDFSZ1,VariantDFSZ2}. The size of flavor-violating axion couplings depends on the corresponding misalignment, i.e. on the unitary matrices that diagonalize SM Yukawas. Thus in the absence of a theory of flavor, these rotations are just described by a bunch of new free parameters, which essentially can be chosen suitably to realize an arbitrary pattern of flavor structures $C^{V,A}_{ij}$. Particular flavor patterns have been employed to e.g.~suppress the axion couplings to nuclei~\cite{Astrophobic1,Astrophobic2} and address stellar cooling anomalies~\cite{SaikawaYanagida}, possibly correlated with low-energy signals from the extra DFSZ Higgs doublet~\cite{MarcinGiovanniMustafa}.     

Particularly motivated and predictive scenarios arise when the PQ symmetry is identified with a flavor symmetry that addresses the SM flavor puzzle, i.e. the hierarchical structure of SM Yukawas~\cite{Wilczek}. For example in the simplest realization the PQ symmetry is identified with a $U(1)_F$ Froggatt-Nielsen symmetry~\cite{FN}, which necessarily has a QCD anomaly~\cite{Anomaly1,Anomaly2}. This allows to predict flavor-violating axion couplings up to model-dependent ${\cal O}(1)$ coefficients~\cite{Axiflavon,Japs}, and determines the phenomenologically most relevant coupling $C^V_{sd}$ to be of the order of the Cabibbo angle, $C^V_{sd} \sim V_{us} \sim 0.1$. Stronger  suppression of light quark transitions arise in models with non-abelian flavor symmetries, for example $U(2)$ models, where $C^V_{sd} \sim V_{td} V_{ts} \sim 10^{-4}$~\cite{U2Axiflavon}.

Finally, even in scenarios with flavor alignment, renormalization group running induces flavor violation~\cite{ChoiMFV, Pospelov, Bauer3} proportional to CKM angles. This leads to strongly suppressed couplings, e.g. $C^V_{sd} \sim y_t^2/(16 \pi^2) V_{td} V_{ts} C_{tt} \log {\Lambda_{UV}/\Lambda_{IR}} \sim 10^{-5}$, which are phenomenologically irrelevant for light axions with masses $m_a \lesssim 100 \, \MeV$, since astrophysical constraints on flavor-diagonal couplings yield much stronger constraints~\cite{KaonReview}.

\section{Precision Flavor Experiments}
Flavor-violating couplings of light DM axions are probed by decays of hadrons and leptons with missing energy.  These signals are very similar to the corresponding SM decays with a final state neutrino pair, which gives an irreducible background that is small in the quark sector, e.g. ${\rm BR} (K \to \pi \nu \overline{\nu})_{\rm SM} \sim 10^{-10}$, and large in the lepton sector, e.g. ${\rm BR} (\mu \to e \nu \overline{\nu})_{\rm SM} = 1$. While baryon and lepton decays are sensitive to both vector and axial vector couplings, parity conservation of strong interactions imply that pseudoscalar meson decays to pseudoscalar (vector) mesons are sensitive only to $C^V$ ($C^A$). In the case of polarized lepton decays one can use the angular distribution of the final lepton to distinguish between vector and axial vector couplings~\cite{LFVaxion} and to effectively suppress SM background, provided that the axion signal is not aligned to it, i.e. $C^A_{ij} \ne - C^V_{ij}$.The axion signal is monochromatic, with an energy of the visible particle close to the kinematic edge of the SM three-body for a light axion\footnote{As discussed above, DM axions must typically be much lighter than an MeV in order to be cosmologically stable. }, i.e. half of the mother particles's mass in the rest frame. While in the lepton sector experiments have been explicitly searching for axionic LFV decays, in the quark sector dedicated searches have been carried out only in the Kaon sector (apart from searches for some flavor-violating $B$-meson decays at CLEO), and there is no published bound on $D \to \pi a$, $B \to K^* a$ and $B \to \rho a$ decays (which are sensitive to $C^V_{cu}, C^A_{bs}, C^A_{bd}$, respectively). However, available information from searches for the SM three-body decays can often be recast to derive bounds on the 2-body decay~\cite{Pospelov}. For example, bounds on $D \to \pi a$ can be obtained from a recast of CLEO data on $D \to (\tau \to \pi \nu) \overline{\nu}$ (as proposed in Ref.~\cite{KamenikSmith}), while bounds on  $B \to K^* a$ can be obtained by recasting BaBar data on $B \to K^* \nu \overline{\nu}$.   
\begin{figure*}[t]
\begin{center}
		\includegraphics[width=1.0\textwidth]{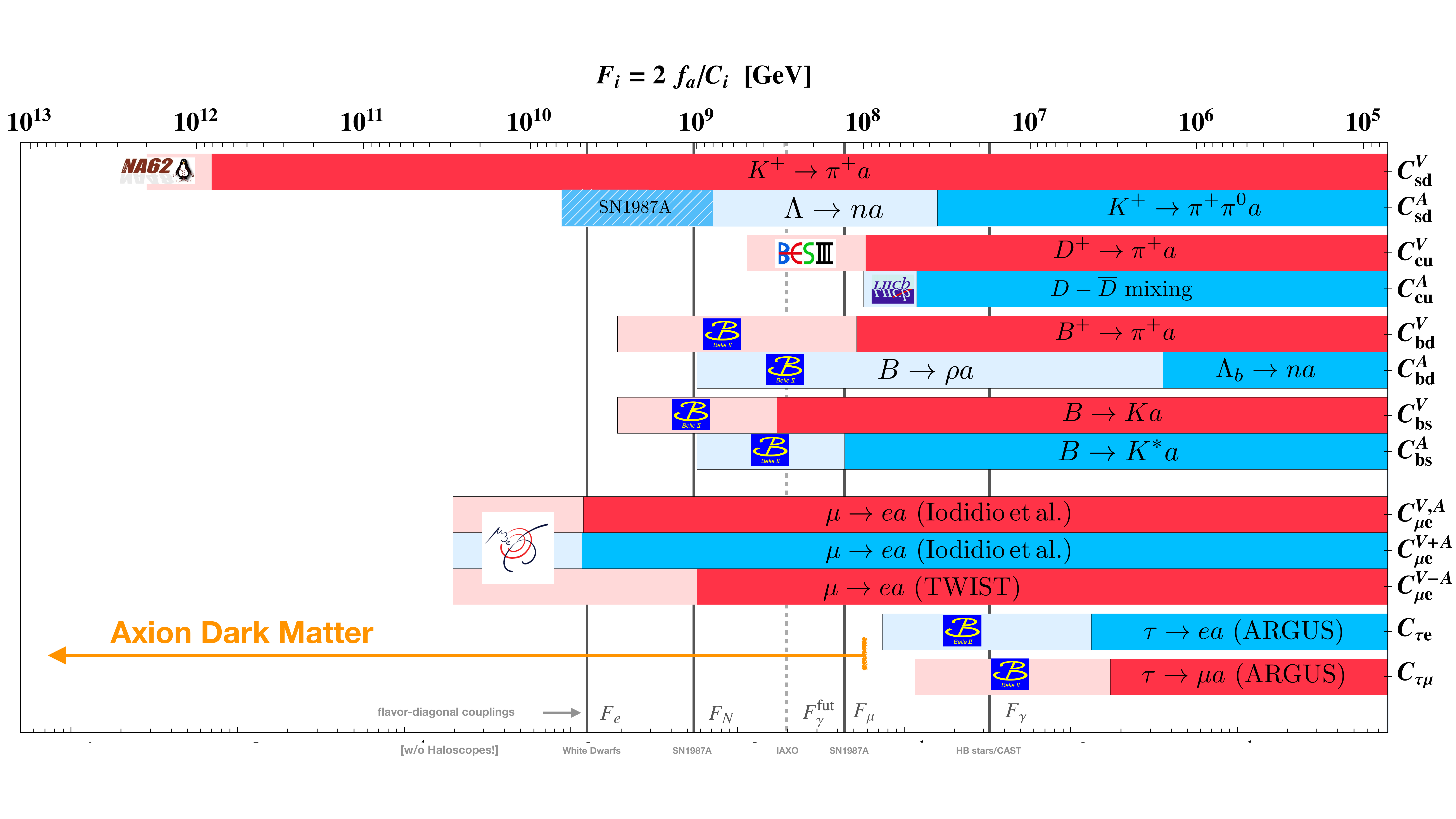}
	\caption{  Present (solid) and future (transparent) constraints on flavor-violating axion couplings. For details on  experiments and observables see Refs.~\cite{Pospelov, LFVaxion}. Also shown is the naive axion DM range, where the axion lifetime exceeds the age of the universe for axion-photon couplings of order unity. The grey vertical lines show the astrophysical bounds on flavor-diagonal couplings $F_i = 2 f_a/C_i$, where $C_i = \{C^A_{ee}, C^A_{NN}, C^A_{\mu \mu} \}$ and $F_\gamma = f_a/C_{a \gamma} = \alpha/(2 \pi g_{a \gamma})$ in the notation of Ref.~\cite{LucaReview}. Not shown are haloscope limits that can exclude QCD axions in a small range close to  $F_\gamma \approx 10^{12} \GeV$~\cite{ADMX3}. \label{Bounds}}
\end{center}
\end{figure*}
A systematic analysis of present and future constraints both in the quark and  lepton sector has been carried out in Refs.~\cite{Pospelov} and \cite{LFVaxion} for a light axion, which is summarized  in Fig.~\ref{Bounds}. The x-axis shows the excluded range of the effective axion coupling $F_i = 2 f_a/C_i$, where $C_i$ is the model-dependent axion coupling shown on the y-axis.  The strongest constraints arise from NA62~\cite{NA62}, which is able to probe scales of the order of $10^{12} \GeV$! Remarkably, also bounds on $\mu \to e a$ are very strong, reaching effective coupling scales of order ${\rm few} \times 10^{9} \GeV$, which are obtained from experiments at TRIUMF in the mid 80's~\cite{Jodidio}. In the future these bounds are expected to be pushed beyond $10^{12} \GeV$ at Mu3e~\cite{Mu3e}, but there are also recent ideas how to reach similar sensitivities at MEG-II~\cite{LFVaxion, DiegoSimon}, which requires precise determination of the SM background close to the endpoint of the Michel spectrum~\cite{Signer}.  Present bounds on all other flavor transitions are weaker, but Belle II has the potential to improve bounds on $B$-meson and $\tau$-decay by orders of magnitudes, reaching scales up to $10^9 \GeV$. 

Finally in the charm sector BES III can perform dedicated searches to establish a bound on $D \to \pi a$, while LHCb will improve bounds on neutral $D$-meson mixing, which at present provides the strongest bound on axial vector couplings to $cu$. Note that in all other flavor transitions constraints from meson mixing are much weaker than the ones from meson decays. This is in contrast to heavy new physics, where constraints on dimension-6 operators from mixing are usually much stronger than constraints from decays, e.g. the bound on the operators $1/\Lambda_{bsbs}^2 (\overline{b} \gamma^\mu s)^2$ and $1/\Lambda_{bs\nu\nu}^2 (\overline{b} \gamma^\mu s)(\overline{\nu} \gamma^\mu \nu)$ are $\Lambda_{bsbs} \gtrsim 10^6 \GeV$ from $B_s$-mixing and $\Lambda_{bs\nu\nu} \gtrsim 10^4 \GeV$ from $B \to K \nu \overline{\nu}$, while the bounds on the dimension-5 operator  $1/F_{bs} \partial_\mu a \overline{b} \gamma^\mu s$ are $F_{bs} \gtrsim 10^8 \GeV$ from the 2-body region of $B \to K \nu \overline{\nu}$ searches. Therefore dedicated searches for 2-body decays with missing energy are extremely interesting, as one can probe much higher scales than by looking for deviations from the SM 3-body decay caused by dimension-6 SMEFT operators, simply because the 2-body decay comes from a dimension-5 operator.

\section{Astrophysics}
While ordinary stars are not hot enough to thermally excite heavy SM flavors, the extreme environment of core-collapse supernovae (SN) features nuclear densities and temperatures of order $40 \, \MeV$, which is sufficient to feature a sizable population of muons and hyperons\footnote{The hyperon-neutron mass difference is $\sim 180 \, \MeV$, resulting in a Boltzmann factor of 1\%}, see e.g. Ref.~\cite{HyperonsSN} for a review. The decay of muons and hyperons to axions  (and electrons/neutrons) then allows for a new source of energy loss in the form of long-lived axions, which is limited by the usual star-cooling constraints on SN1987A, giving a rough upper bound on the energy loss rate per unit mass of $\epsilon_a \lesssim 10^{19} {\rm erg/(g \, sec )}$. This rate can be readily calculated, and a back-of-the-envelope estimate gives for the energy loss rate per unit volume ~\cite{Pospelov}
\begin{align}
\rho \epsilon_a \sim  n_n  e^{-(m_\Lambda - m_n)/T} (m_\Lambda - m_n) \Gamma (\Lambda \to n a) \, , 
\end{align}
where $n_n$ is the number density of neutrons and $\rho$ is the mass density of the proto-neutron star. Taking nuclear densities for simplicity, this leads to a upper bound on the branching fraction ${\rm BR}(\Lambda \to n a)$ of order $10^{-8}$, which is about 6 orders of magnitude better than the bound one can infer from observed hyperon decays and the total hyperon width~\cite{Pospelov}. This very rough estimate is confirmed by a more refined calculation carried out in Ref.~\cite{Jorge}, using the full radial profiles from state-of-the-art SN simulations combined with a hyperonic equation of state. Also including medium effects in the decay rate, one obtains a bound  ${\rm BR}(\Lambda \to n a) \lesssim 8.0 \times 10^{-9}$, which is valid for any sufficiently long-lived and free-streaming particle $a$. This bound can readily converted to the strongest constraint on the axial vector coupling $2 f_a/C^A_{sd} \gtrsim {\rm few} \times 10^{9} \GeV$, which is of the same order as the bound on electron couplings from white dwarf cooling. 

In a similar way one can calculate the SN1987A constraint on axion production from muon decays~\cite{LFVaxion}, giving a bound on  the branching ratio of order ${\rm BR}(\mu \to e a)  \lesssim {\rm few} \times 10^{-3}$, which is however more than two orders of magnitude below the laboratory bound, even for $V-A$ axion couplings aligned to the SM weak decay~\cite{TWIST}.

\section{Cosmology}
Flavor-violating meson or lepton decays can produce axions also in the early universe. If these axions are very light, such as the QCD axion, this would give rise to a thermal population of axions that is strongly constrained by CMB bounds on Dark Radiation, usually formulated as a bound on the effective number of new neutrino species $\Delta N_{\rm eff}$. These constraints have been studied in detail in Ref.~\cite{Francesco}, finding that future cosmological
bounds from CMB-S4 surveys will often be comparable or even stronger than the ones obtained in laboratories, which gives additional motivation to devise and optimize dedicated axion searches at Belle II and BES III\footnote{For $d-s$ and $\mu - e$ transitions present laboratory bounds are much stronger than prospective CMB-S4 bounds.}. 

For heavy axions the $\Delta N_{\rm eff}$ bounds can be avoided, and one can consider scenarios where the flavor-violating decays are in fact the dominant production of axion DM~\cite{LFVFreezein}, reproducing the observed DM abundance via thermal freeze-in~\cite{FreezeIn}. Since in this case the relic abundance is set by the product of axion mass and flavor-violating decay rate, $\Omega_a h^2 \propto m_a \Gamma (f_i \to f_j a) \propto m_a (C_{ij}/f_a)^2$, this gives a prediction for the decay as a function of the axion mass, which otherwise requires extensive flavor model-building. It is clear that this can work only for axion masses in a suitable range, which is bounded from above by requiring that the decay is kinematically allowed, and a lower bound because there is an experimental upper bound on the decay rate, besides constraints from warm Dark Matter (WDM). 

The main challenge of this scenario is DM stability, as the axion can always decay into two photons, with a decay rate set by 
\begin{align}
\Gamma_{a \to \gamma \gamma} &\approx \frac{\alpha_{\rm em}^2}{64 \pi^3} \frac{m_a^3}{f_a^2}  \left| E- 1.92 N +\sum_i C^A_{ii} \frac{m_a^2}{12 m_{i}^2}\right|^2 \, , 
  \end{align}
  up to higher powers of $m_a^2/m_i^2$, and the sum runs over all SM fermions and $N$ ($E$) is the color (electromagnetic) anomaly coefficient. While it is relatively easy to make the axion lifetime larger than the age of the universe, $1/H_0 \sim 10^{17} {\rm sec}$, it turns out that in the relevant axion mass range much stronger constraints on the decay rate arise from X-ray telescopes, requiring $\Gamma_{a \to \gamma \gamma} \gtrsim 10^{28} {\rm sec}$~\cite{LFVFreezein}. This implies that the underlying PQ symmetry must be anomaly-free, and one needs $m_a \ll m_i$ and/or $C_{ii} \ll C_{ij}$. Note in fact that DM production is controlled by $C_{i \ne j}$ while DM stability is governed by $C_{ii}$, so in principle these are independent parameters, although a strong hierarchy  $C_{ii} \ll C_{ij}$ seems somewhat unnatural.  
  
Simple realizations of this idea have been constructed for the lepton sector in Ref.~\cite{LFVFreezein}. Restricting to an effective 2-flavor scenario, where two right-handed SM leptons are oppositely charged under PQ and a single angle parametrizes the rotation to mass basis, one has just three parameters: the axion mass $m_a$, the axion decay constant $f_a$, and a single angle $0 \le \alpha \le \pi/2$, which controls the couplings to leptons, e.g. for the $\mu e$ scenario $C_{ee} = - C_{\mu \mu} = \sin \alpha$, $C_{ \mu e} = C_{e \mu } = \cos \alpha$. One of these parameters is fixed by requiring that $\mu \to e a$ decays gives the dominant contribution to the axion relic abundance via freeze-in, which one can choose to be $f_a$. Contributions from other thermal processes such as $h \mu \to e a$ are UV-sensitive~\cite{FreezeIn}, and are sub-dominant for sufficiently low reheating temperatures $T_R \lesssim 10^7 \GeV$. A non-thermal contribution to the abundance comes from misalignment, which is however not the usual one for the QCD axion, as the relevant axion masses of interest ($m_a \gg {\rm keV}$ from the WDM bound) and the lower bound on the reheating temperature imply that coherent axion oscillations start prior to reheating. For an initial period of matter domination, the present day axion abundance from misalignment  is then suppressed by the dilution which occurred between the onset of oscillations and the begin of radiation domination, given by a factor $T_R/T_{\rm osc}^{\rm std}$, where  $T_{\rm osc}^{\rm std} \propto \sqrt{m_a M_{\rm Pl}}$ is the standard oscillation temperature. 
\begin{figure*}[t]
\begin{center}
		\includegraphics[width=1.0\textwidth]{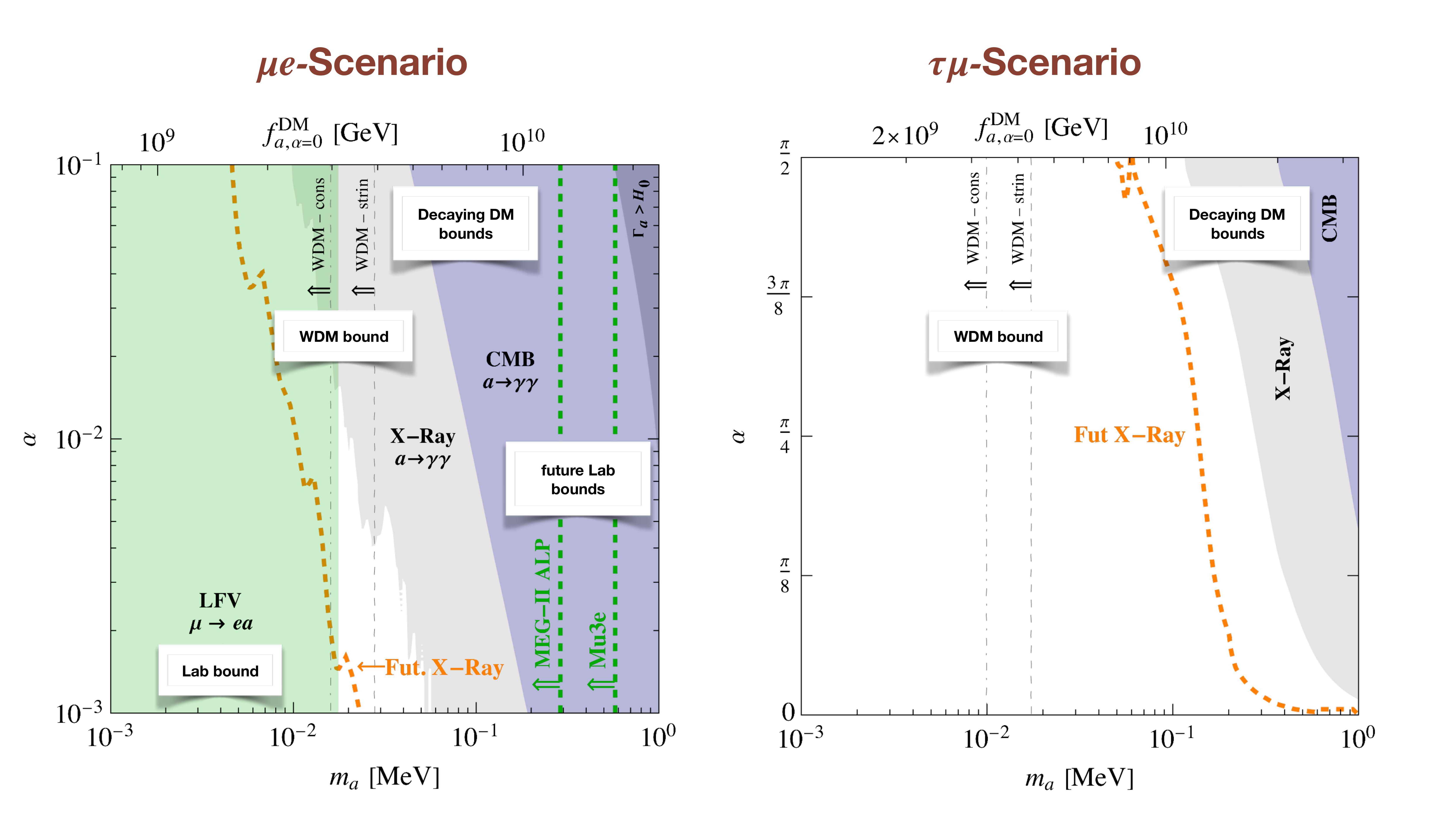}
	\caption{ Allowed parameter space for DM freeze-in through LFV decays for the $\mu e$- (left panel) and $\tau \mu$- (right panel) scenarios. The decay constant $f_a$ (top x-axis) is determined by requiring that the DM abundance today is produced through freeze-in, once the ALP mass $m_a$ (bottom x-axis) and the mixing angle  $\alpha$ (y-axis) is fixed to the reference value $\alpha  = 0$. The dark blue shaded, blue shaded and gray shaded regions are excluded by the DM lifetime, CMB and X-ray constraints on decaying DM, respectively, while the reach of future X-rays searches is shown by dashed orange lines. Conservative (stringent) constraints on WDM are shown as dotted-dashed (dashed) gray lines. The present bound from searches for $\mu \to e a$ are shown as green shaded regions, while the prospects for future proposed searches at MEG II and Mu3e are shown as dashed green lines. For $\tau \to \mu a $  experimental prospects are limited to $f_a \lesssim 10^8 \GeV$, and thus not visible in the figure.  \label{Results}}
\end{center}
\end{figure*}

Numerical results for the $\mu e$-Scenario and the $\tau \mu$-Scenario (the $\tau e$-Scenario is very similar to this) are shown in Fig.~\ref{Results}. After fixing the axion decay constant to reproduce the observed abundance, one is left with a two-dimensional parameter space, where constraints on decaying DM (age of the universe, CMB and X-ray telescopes), Warm DM and direct laboratory searches (see Section 3) are imposed. For the $\mu e$-Scenario the stringent bounds from X-ray searches require a sizable hierarchy between off-diagonal and diagonal axion couplings of at least 1\%, in particular ruling out the limit of exact flavor conservation $\alpha = \pi/2$. Present limits from laboratory searches are roughly at the same level as WDM constraints, but proposed searches at
MEG II and Mu3e will probe almost the entire allowed parameter space. In contrast experimental prospects for Belle 2 are not sensitive to test the viable parameter space of the $\mu e$-Scenario, which is constrained from below by the WDM bound and from above by X-ray searches. Constraints on decaying DM are weaker than in the $\mu e$-Scenario, because of the suppression of $\Gamma_{a \to \gamma \gamma}$ by the muon instead of the electron mass, resulting in an additional suppression factor $(m_e/m_\mu)^4$ in the decay rate. This ensures DM stability even for an ${\cal O}(1)$ hierarchy between flavor diagonal and off-diagonal couplings, allowing a large viable parameter region that will be probed by future X-ray telescopes.

\section{Summary and Conclusions}
To summarize, I have argued that invisible axions with flavor-violating couplings can be efficiently produced from decays of SM particles, which is relevant for precision flavor experiments, astrophysics and cosmology. 
\begin{itemize}
\item First, flavor-violating decays with final states axions can be searched for at colliders, where they mimic the corresponding SM 3-body decay with a neutrino pair, but with a visible monochromatic daughter particle. Explicitly looking for such decays allows to probe enormously large scales, since in contrast to NP contributions to the SM 3-body decay, the axionic 2-body decay is controlled by dimension-5 operators. Specifically, NA62 will probe scales of order $10^{12} \GeV$, LFV experiments such as Mu3e or MEG II scales of order $10^{10} \GeV$, and Belle II can substantially improve existing constraints for $b$-quark transitions, testing scales of order $10^{9} \GeV$ (see Fig.~\ref{Plane} for a comparison with the usual axion plane).  
\begin{figure*}[t]
\begin{center}
		\includegraphics[width=1.0\textwidth]{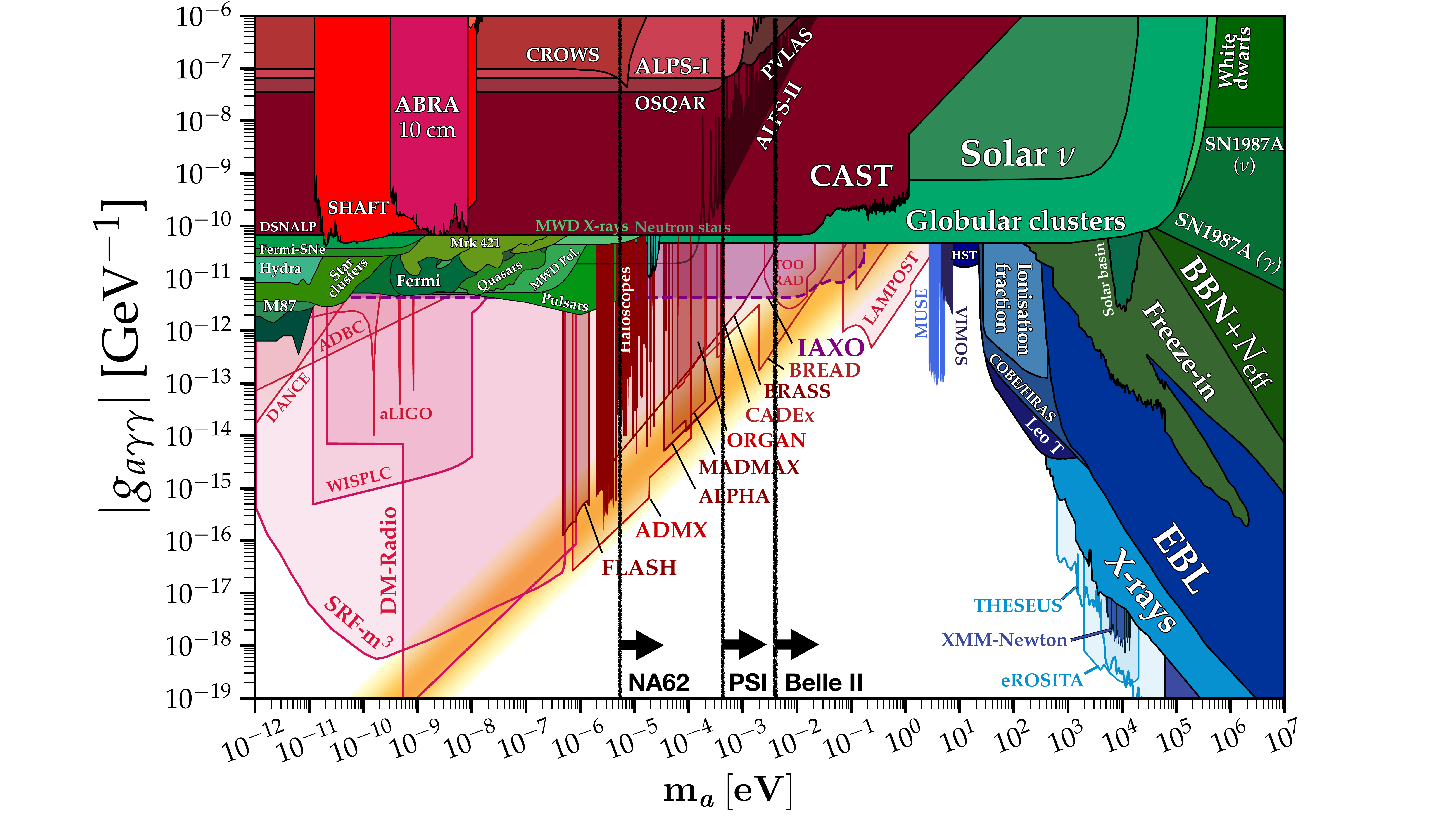}
	\caption{ Selected prospects from precision flavor experiments in the $m_a-g_{a \gamma \gamma}$ plane., probing $K^+ \to \pi^+ a$ (NA62), $\mu^+ \to e^+ a$ (PSI = Mu3e/MEG II) and $B^+ \to K^+/\pi^+ a$ (Belle II). The bounds have been obtained using the results of Ref.~\cite{Pospelov} with unit flavor-violating couplings $C_{ij} = 1$, and converted to an upper bound on $m_a$ using the QCD axion relation.   Non-flavor bounds are taken from \href{https://cajohare.github.io/AxionLimits}{https://cajohare.github.io/AxionLimits}. \label{Plane}}
\end{center}
\end{figure*}
\item In the extreme environments of core-collapse Supernovae moderately heavy flavors are thermally excited, and a sizable population of muons and hyperons are present in the dense proto-neutron star formed few seconds after core-collapse. Their decays to invisible axions provides an efficient contribution to energy loss in form of axions, which is bounded from above by the observed neutrinos in the burst of SN1987A. While the constraint on muon decays is weaker than the laboratory bound, the constraint on the  $\Lambda \to n a$ branching ratio is of order $10^{-8}$, which exceeds terrestrial limits by many orders of magnitude.   
\item Flavor-violating decays will take place in the early universe, providing an important source of axion production. While the decays of light DM axions are bounded from their contribution to the effective number of relativistic degrees of freedom, heavy axions in the keV - MeV range can be produced in the right amount  to fully account for the observed DM abundance. For the case of production from LFV decays, this gives rise to very simple DM models, which are  mainly constrained by X-ray searches for decaying DM. For the case of muon decays, much of the allowed parameter space will be probed by searching for $\mu \to e a$ at PSI, probing the very same decay that could have been responsible for DM production in the early universe. 
\end{itemize}
\section*{Acknowledgements}
I would like to thank my collaborators Lorenzo Calibbi, Jorge Martin Camalich, Paolo Panci, Maxim Pospelov, Diego Redigolo, Thomas Schwetz, Jorge Terol-Calvo, Laura Tolos, Pham Ngoc Hoa Vuong and Jure Zupan.
\bibliographystyle{JHEP}
\bibliography{MyBib}
\end{document}